\documentstyle[11pt,aaspp4]{article} 

\begin{document}



\def\etc {{\it etc.}\ }
\def\eg {{\it e.g.}\ }
\def\ie {{\it i.e.}\ }
\def\etal {{\it et al.}\ }
\def\cf {{\it cf.}\ }
\def\percent{{\it per cent}\ }
\def\spose#1{\hbox to 0pt{#1\hss}}
\def\lta{\mathrel{\spose{\lower 3pt\hbox{$\mathchar"218$}}
     \raise 2.0pt\hbox{$\mathchar"13C$}}}
\def\gta{\mathrel{\spose{\lower 3pt\hbox{$\mathchar"218$}}
     \raise 2.0pt\hbox{$\mathchar"13E$}}}

\title{The Luminosity Distribution of Local Group Galaxies} 
\author {Christopher J. Pritchet}
\affil {Department of Physics \& Astronomy, University of Victoria,
  P. O. Box 3055, Victoria, British Columbia, V8W 3P6, Canada}
\author {Sidney van den Bergh}
\affil {Dominion Astrophysical Observatory, Herzberg Institute of
  Astrophysics, National Research Council, Victoria, British Columbia,
  V8X 4M6, Canada} 

\begin{abstract} 

From a rediscussion of Local Group membership, and of distances to
individual galaxies, we obtain $M_V$ values for 35 probable and possible
Local Group members. The luminosity function of these objects is well
fitted by a Schechter function with faint end slope $\alpha = -1.1 \pm
0.1$. The probability that the luminosity distribution of the Local
Group is a single Schechter function with $\alpha$ steeper than $-1.3$
is less than 1 per cent.  However, more complicated luminosity functions,
such as multi-component Schechter functions with steep faint-end
slopes, cannot be ruled out. There is some evidence that the luminosity
distribution of dwarf spheroidal galaxies in the Local Group is steeper than that of dwarf irregular galaxies.

\end{abstract}

\keywords{galaxies: Local Group --- galaxies: luminosity function}

\section{Introduction}

The galaxy luminosity distribution, or luminosity function\footnote{ The
{\it luminosity function} is expressed in units of number density (\eg
number per Mpc$^3$), whereas the {\it luminosity distribution} simply
gives the shape of the luminosity function without density normalization.},
$\phi(L)$  plays
an important role in our understanding of the properties of galaxies,
galaxy evolution, and galaxy formation. The connection between $\phi(L)$
and galaxy formation is through the primordial
density fluctuation spectrum, $\delta\rho/\rho \propto k^n$, where
$k$ is wavenumber. If the Universe consisted only of weakly interacting
particles (\eg ``cold dark matter''), then the mass function of ``halos''
would be determined solely by $n$, and could be computed using a simple
physical recipe for the gravitational clustering and merging of
halos (\eg Press \& Schechter 1974). However, it is the baryons,
rather than the dark matter, that we observe directly; hence
the luminosity function $\phi(L)$ depends on gas physics and
radiation processes (\eg cooling, radiative transfer, star formation, 
energy input from supernovae, to name just a few). It follows that
the luminosity function
is sensitive not only to the fluctuation spectrum $\delta\rho/\rho$, but also
to the detailed history of galaxy formation and evolution in different
environments.

Generally the luminosity function of galaxies is parametrized by a Schechter
(1976) function,

\begin{equation}
\phi(L) = \phi^* e^{-(L/L^*)} (L/L^*)^{\alpha},
\end{equation}

\noindent
where $L^*$ is a characteristic luminosity defining the transition
between a power-law at faint magnitudes and an exponential cutoff at
bright magnitudes. [Further information concerning the Schechter function
can be found in Felten (1985).] $L^*$ corresponds roughly to the brightness
of the Milky Way. For a CDM fluctuation spectrum with
$n=1$, the power-law exponent $\alpha$ is theoretically predicted to be
$\approx -2$ on the scale of galaxies (Bardeen \etal 1986), though this
result is very sensitive to the detailed physical processes involved in
the calculation (\cf Babul and Ferguson 1996, Frenk \etal 1996, Kauffmann
\etal 1998).

What is known empirically about the shape of the luminosity distribution in
the nearby Universe?  Values of $\alpha$ in the range $-$0.7 to $-$1.0 have
been obtained for bright galaxies within 2--3 magnitudes of $L^*$ (\eg
Loveday \etal 1992; Marzke \etal 1994 $a,b$; Lin \etal 1996), with a hint
of a turnup in the LF at magnitudes fainter than about $M_R = -17$ in the
work of Lin {\it et al.}. The field luminosity distribution derived from the SSRS2
redshift survey (Marzke \etal 1998) indicates a relatively flat value of
$\alpha \simeq -1$ for E/S0's (including dwarf spheroidals) and spirals,
and a steep $\alpha = -1.8$ value for dwarf irregulars and peculiars; this
steepening of the LF for late-type star forming systems also appears in
the work of Bromley \etal (1998), who subdivided the Las Campanas Redshift
Survey according to emission-line strength.  C\^ot\'e \etal (1998) have
found a very steep ($\alpha=-2.1$) luminosity distribution for
nearby H~I-rich low surface brightness galaxies, and Schneider, Spitzak,
\& Rosenberg (1998) find a steep upturn in the H~I mass function for
low mass objects (M$_{H~I} \lta 10^8$ M$_\odot$). 

Turning to other environments, Loveday (1997) finds that the luminosity
distribution of dwarf galaxies surrounding luminous ($\sim L^*$)
galaxies is steep: this can be interpreted either as a turnup in a
supposedly universal field luminosity distribution, or alternately as
an enhanced probability that dwarfs form in the vicinity of luminous,
massive galaxies. Galaxies in groups exhibit a slope $\alpha \simeq
-1$ (Muriel, Valotto \& Lambas 1998). There is evidence that compact
group luminosity distributions cannot be fitted by a single Schechter
function (Hunsberger, Charlton \& Zaritsky 1998), but instead show
$\alpha =-0.5$ at the bright end and $\alpha=-1.2$ below $M_R \simeq
-16$. There is also some indication that cluster LF's must be fitted
with multiple Schechter functions (\eg Trentham 1998$a$, Lopez-Cruz
\etal 1997), with a steep upturn at faint magnitudes (\eg Trentham
1998$b$; Phillipps \etal 1998$a$). Phillipps \etal (1998$b$) suggest
the existence of an environmental dependence of dwarf-to-giant ratio
(\ie $\alpha$) in clusters.  The faintest galaxies in the steep
luminosity distribution population in clusters are, based on their colors
(Trentham 1998$a,b$), dwarf spheroidals, whereas in the field they appear
to be gas-rich dwarf irregulars (Marzke \etal 1998). (It should, however, be noted
that the Marzke \etal dIr/peculiar sample is drawn from a small local
volume and comprises only 4\% of the total SSRS2 sample.)
 
On the basis of the discussion given above, it appears that
the simple paradigm of a universal Schechter function (that fits
the LF of galaxies in all environments) is now untenable.  There is
growing evidence that the LF is not a simple Schechter function, that it
depends on environment, and that it also depends on galaxy morphological
class and/or gas content within a given environment.  No clear physical
picture has emerged that would allow one to understand current observational
evidence on the shape and environmental dependence of the LF.

The Local Group represents a unique opportunity for the study
of the luminosity distribution in relatively low density
environments. The manner in which faint
Local Group galaxies are detected is completely different from that for
other groups (because most Local Group galaxies are easily resolved into
stars); hence surface brightness selection effects operate differently
in the Local Group than they do in more distant groups. Subtraction of
foreground and background contaminating objects is irrelevant for Local
Group galaxies, something that is of course not the case for more distant
clusters. The numbers of galaxies in poor groups are so low that it has
usually been possible to study only the {\it composite} LF of poor groups
(rather than the LF of any group individually) -- here the Local Group
again is an exception. Furthermore, the census of Local Group members
extends to considerably fainter absolute magnitudes ($M_V \simeq -8.5$) than
do any other samples for which the luminosity distribution has been
measured (though incompleteness must be severe at the faint end). Thus a study of the
Local Group luminosity distribution is of considerable importance.

What is known about the Local Group luminosity distribution? Tully (1988)
derived a composite luminosity function for six nearby groups (including
the Local Group), and found $\alpha=-1\pm0.2$. Van den Bergh
(1992) demonstrated that the integral luminosity function of the Local
Group was consistent with $\alpha = -1.1$, but did not rule out the
possibility that other values of $\alpha$ fitted the data equally
well. More recently, Mateo (1998) has shown that the LF of galaxies in
the vicinity of the Local Group (but extending out beyond the usually
accepted LG boundary of R = 1 Mpc) is consistent with that derived for
poor groups (Ferguson and Sandage 1991). Again, this statement does not
preclude the possibility that other luminosity distributions fit equally 
well.

Over the past few years substantial additional data have been accumulated
on Local Group membership and absolute magnitudes, and so the time seems
ripe for a fresh, and more detailed, attack on the problem of the Local
Group luminosity distribution.

\section {The Local Group Catalog}

   The Local Group of galaxies was first decribed by Hubble (1936) in
his book {\it The Realm of the Nebulae}. He listed M~31, M~32, M~33, the
Magellanic Clouds, NGC~205, NGC~6822, and IC~1613 as probable members
of the small group of galaxies associated with our Milky Way system.
Inspection of the prints of the Palomar Sky Survey (van den Bergh 1962)
shows that a large fraction of all of galaxies occur in small groups
and clusters that resemble the Local Group. This shows that our Galaxy
is located in a rather typical region of space. Since Hubble's
pioneering work the number of galaxies that are known to belong to the
Local Group has increased by 4 or 5 per decade to over thirty. A
listing of data on presently known Local Group members (van den Bergh
2000) is given in Table 1.

Selection of Local Group members proceeded in three steps. First
galaxies with distances from the Local Group centroid (Courteau and
van den Bergh 1999) less than or about 1.5~Mpc were regarded as suspected Group
members. Secondly it was required that Local Group members should lie
close to the relation between between radial velocity V$_r$ and cos
$\theta$   for well-established Local Group members, where $\theta$
is the distance from the solar apex (Courteau and van den Bergh 1999).
Finally, Local Group members should not appear to be associated with
groups of galaxies that are centered well beyond the limits of the
Local Group.  

On the basis of these criteria van den Bergh (1994, 2000)
concluded that the following objects should be excluded from Local
Group membership: (1) UKS~2323-326, (2) Maffei~1 and its companions,
(3) UGC~A86, (4) NGC~1560, (5) NGC~1569, (6) NGC~5237, (7) DDO~187, (8)
Cassiopeia~1, and (9) NGC~55. A particularly strong concentration of
these Local Group suspects, which includes (2), (3), (4), (5) and (8)
listed above, occurs in the direction of the IC~342/Maffei group
(Krismer, Tully \& Gioia 1995). (1) and (9) appear in the
direction of the Sculptor (=South Polar) group; in the case of (9),
Jergen, Freeman \& Binggeli (1998) find D=$1.66 \pm 0.2$ kpc, which gives
a distance 1.65 Mpc from the Local Group centroid. Finally, the discovery
of a Cepheid (Tolstoy et al. 1995) in DDO~155 (=GR~8) suggests that
this object is located at a distance of 2.2 Mpc, which places it well
beyond the usually accepted limits of the Local Group. Dohm-Palmer \etal
(1998) obtain a similar distance to DDO~155 from the tip of the red giant 
branch.

Also excluded
from Local Group membership are the galaxies NGC~3109, Antlia, Sextans~A
and Sextans~B. These objects, which are located relatively close
together on the sky, all have distances of 1.3 -- 1.5 Mpc from the
Milky Way, and, of more relevance, distances of $\sim 1.7$~Mpc from
the Local Group centroid. Furthermore, these objects possess a mean
radial velocity of $+114 \pm 12$ km s$^{-1}$
relative to the relation between $V_r$ and cos $\theta$ for
well-established Local Group members (van den Bergh 1999).
This suggests that these galaxies form a small group just beyond the
zero velocity surface of the Local Group. (This surface is at a distance
$R(LG) = 1.18 \pm 0.15$ Mpc from the Local Group centroid
[Courteau and van den Bergh 1999].)

How does the Local Group membership defined above compare with that of
Mateo (1998)? The principal differences are that the Mateo catalog does
not contain several recently-discovered satellites of M~31, but does
include nine objects beyond 1 Mpc (NGC~55, EGB~0427+63, Sextans A,
Sextans B, NGC~3109, Antlia, GR~8, IC~5152, UKS~2323-326). Most of these
were discussed above. EGB~0427+63 has a distance of 2.2 Mpc
(Karachentsev, Tikhonov \& Sazonova 1994) and thus lies well outside
the Local Group. From a color-magnitude diagram, Zijlstra \& Minniti (1999) find that IC~5152 has a distance from the Milky Way of $1.70 \pm 0.16$ Mpc,
a result that agrees with the Cepheid distance of 1.6 Mpc (Caldwell \&
Schommer 1988); the distance of this galaxy from the Local Group centroid
is therefore 1.8 Mpc, again beyond the Local Group.

A more detailed discussion of
Local Group membership and of the outer boundary of the Local Group can be
found in van den Bergh (2000).

\section{Luminosity Distribution of the Local Group}

   Because Local Group galaxies are situated so nearby it is possible to
study their luminosity distribution down to very faint absolute
magnitudes. Nevertheless these data are, no doubt, still quite
incomplete for M$_V > -10$. This is shown most clearly by the fact that only one
galaxy fainter than this limit has so far been discovered in the
Andromeda subgroup of the Local Group, whereas five such faint objects
are presently known in the Milky Way subgroup of the Local Group. 
On the other hand, a
survey of a twenty thousand square degree area at high Galactic
latitudes by Irwin (1994) resulted in the discovery of only a single
new Local Group member. Furthermore, no new optically visible Local Group
galaxies have turned up in the survey of compact high latitude
high velocity clouds (Braun and Burton 1999).
Taken at face value these results might be taken to suggest
that the luminosity distribution of the Local Group no longer increases
below $M_V \simeq -8$. It is noted in passing that very large low surface
brightness galaxies in the Local Group, like those that have been discovered in the Virgo
cluster (Impey, Bothun \& Malin 1988), in the Fornax cluster (Bothun,
Impey \& Malin 1991) and in the M~81 group (Caldwell et al. 1998), may
have also eluded us.

The data in Table 1 can be used to study the luminosity
distribution of Local Group galaxies. Histograms plotting this distribution
are shown in Figure 1. The upper histogram clearly shows an increased
number of objects at faint absolute magnitudes. The separation by
morphological type (lower two panels) shows that most of this increase
is due to galaxies that are dwarf spheroidals.

A somewhat smoother visual impression of the Local Group luminosity
distribution may be obtained by plotting the {\it cumulative} luminosity
distribution, which is compared in Figure 2 to several different
cumulative Schechter functions.  In Fig. 2, the cumulative numbers are
normalized at M$_V$=$-$10 because it is unlikely that the data are complete
at fainter magnitudes.  This figure shows that a Schechter function with
$\alpha\simeq -1.1$ and M$_V^*=-20$ is an acceptable fit to the data (\cf van den Bergh
1992, Mateo 1998), and that there is some evidence for a steepening to
$\alpha < -1.3$ at faint magnitudes.  (Because of incompleteness effects,
this is an upper limit on the faint-end slope).

To parametrize the luminosity distribution of the Local Group, we fit
the data from Table 1 to a Schechter (1976) function [Eqn. (1)]. As
discussed in \S 1, this function possesses a power-law luminosity
dependence with exponent $\alpha$, and an exponential cut-off at
L$>$L$^*$.  In a plot of log $\phi$(M) vs. absolute magnitude M, the faint end of
the Schechter function is linear, with slope $a = -0.4 (\alpha + 1)$.
In detail, we fit the unbinned Local Group absolute magnitude data to a
Schechter function using maximum likelihood techniques. The small
number of objects involved dictates that we use a Poisson,
rather than Gaussian, estimator of likelihood. The very small population of luminous galaxies ($L \gta L^*$) also means that it is not possible to obtain 
a robust estimation of M$_V^*$; hence we
fit only $\alpha$ (and of course a normalization constant proportional to
$\phi^*$), with a few different trial values of M$_V^*$ (which make virtually
{\it no} difference to the fitted value of $\alpha$). The sparsenesss
of the dataset furthermore prevents us from considering luminosity
distributions that are a combination of two or more fitting functions (as
found for Coma and other rich clusters by Trentham 1998$b$; see also
Ferguson and Sandage 1991; Binggeli, Sandage and Tammann 1987). However,
as will be seen, it is nevertheless possible to constrain such functions
by isolating different magnitude ranges.

The maximum likelihood program was tested with artificial data sets drawn
from a distribution of absolute magnitudes that followed a Schechter
function. From thousands of simulations, the maximum likelihood program
was found to return almost precisely the input value of $\alpha$ in the
mean, even for very small numbers of objects (N $<$ 10). Furthermore,
the error estimates (see below) were also found to be accurate.

Table 2 gives the maximum likelihood value of $\alpha$ for various fits
to the Local Group data (different magnitude ranges, and selections of
morphological types), together with several different error estimators
for this quantity. The first error estimate is simply a $1\sigma$ error:
this is derived by finding that region of the maximum likelihood
probability distribution
that is centred on the fitted value of $\alpha$, and that includes 68\%
of the probability. Also given in Table 2 are 95\% and 99\% confidence
limits for the upper bound on $\alpha$ (\ie the values of $\alpha$ for
which the probability is 95\% and  99\% that $\alpha$ is steeper than
this). Finally, Table 2 also shows the percentage probability that
$\alpha$ is steeper than (less than) --1.3.

Considering the entire data set (faint limit $M_V = -8$), it is
apparent that a value $\alpha=-1.07 \pm 0.05$ is a best fit, with a 95\%
probability that $\alpha < -0.98$. Since the Local Group luminosity
distribution is known to be incomplete at such faint magnitudes,
we instead consider limiting the choice of objects at the faint end.
However, regardless of the choice of parameters, the value $\alpha \simeq
-1.1 \pm 0.1$ emerges. The probability that $\alpha$ is steeper
than --1.3 is $<1$\%. 

The only exception to this is for galaxies fainter than $M_V = -15$; for
such objects slopes as steep as $\alpha=-1.5$ are derived.  However: (1)
these slopes have large errors ($\pm 0.2$ or even greater) because they
are based both on small numbers of objects, and also on a limited range
of $M_V$; (2) the 95\% probability upper limit for $\alpha$ (lower
limit on steepness) continues to hover around $\alpha = -1$, and the
probability that $\alpha < -1.3$ is only 85\%; and (3) the effect goes
away if one instead considers objects brighter than $M_V = -16$. Thus
we consider this apparent steepening of the luminosity distribution at
faint absolute magnitudes to be tantalizing, but {\it not} proven.

Note that the derived slope is not very sensitive to the precise value of
the faint cutoff for the fit. This is probably because of incompleteness
at the faint end, but also because, even with a faint end upturn in the
luminosity distribution, the majority of objects contributing to the
fit are at brighter magnitudes.

The derived value of the slope $\alpha$ is not sensitive to the assumed
outer boundary of the Local Group. Relative to the Local Group centroid,
the shell between 1.18 Mpc (the zero velocity surface according to
Courteau and van den Bergh 1999) and 1.6~Mpc contains only a single
galaxy, SagDIG. Removing this galaxy from our sample does not alter
any of the results above. Mateo (1998) includes nine galaxies in
his Local Group catalog that, because of their distance, do not appear
in our catalog (see Section 2). Including these nine objects in our fits makes
$\alpha$ less steep by $< 0.1$. It should be noted that even this small
effect can be entirely explained by incompleteness at the low luminosity
end of the sample of galaxies beyond 1.18 Mpc. We also stress that
the available evidence does {\it not} support the inclusion of these nine
galaxies in our Local Group catalog (see discussion in Section 2).

Most of the apparent steepening in $\alpha$ for faint objects is due to
the dwarf spheroidals in the Local Group, as can be seen from Fig. 1.
Fitting a power-law slope to these objects alone shows a steeper
value of $\alpha$ than for the entire dataset, but again the effect is only
marginally significant. From a Kolmogorov-Smirnov two-sample test,
the difference between the luminosity distributions of dIr's and
dSph's is significant only at the 90\% level. Unfortunately, the observations
of M81 group dwarfs (\eg van Driel \etal 1998) do not enable us to throw
additional light on this problem. This is because these authors were
not able to determine morphological classes for the two faintest magnitude
bins in their survey.

Trentham (1998$c$) has derived a composite luminosity function for
galaxies in clusters, and has shown that it can be applied to galaxies
in the field as well. The cumulative form of this empirical luminosity
function (for which $\alpha$ steepens towards fainter $M_V$) is plotted as
the dashed line in Fig. 2. A Kolmogorov-Smirnov test excludes the
possibility that Local Group galaxies are drawn from this parent
population at $>99$\% probability. 

Finally, we have compared the distribution of $B$ magnitudes of galaxies
in the M~81 group (van Driel \etal 1998) with the luminosity distribution
of M$_V$ of Local Group galaxies, under the assumptions that (m-M)$_B =
28.8$ and $<B-V> = 0.5$ for the M~81 galaxies. From a comparison between
the (presumably more-or-less complete) data on galaxies with M$_V$
brighter than --10 and $B$ brighter than 17.5, a Kolmogorov-Smirnov
test shows no significant difference between the M~81 (N=38) and Local Group
luminosity distributions (N=27). This suggests that the Local Group and M~81
LFs are broadly similar, and are drawn from similar parent populations.

\section {Discussion and Conclusions}

A Schechter function with $\alpha \simeq -1.1 \pm 0.1$ provides a good
fit to new data for the luminosity distribution of the Local Group.
This result is in agreement with the luminosity distribution found
for poor groups (\eg Ferguson and Sandage 1991, Muriel \etal 1998),
and is probably consistent with the work of Hunsberger \etal (1998), who
found $\alpha \simeq -0.5$ for $M_R < -18$ and $-$1.2 for $M_R > -18$.
Our result is comparable to various determinations of $\alpha$ in the
field (\eg Loveday \etal 1992; Marzke \etal 1994 $a,b$; Lin \etal 1996;
Marzke \etal 1998), and is insensitive to the manner in which the
Local Group is defined.

There is evidence for a steepening in $\alpha$ below M$_V$=$-$15; as
discussed in \S 1, this effect has been observed in other environments.
However, the steepening of the field luminosity distribution observed by
Marzke \etal (1998) is in the dIr population, in contrast to the
situation in the Local Group,
for which the dIr population possesses a flat $\alpha \approx -1$, and
for which the dSph population appears to be possess steeper $\alpha$
(though this difference is significant only at the $\sim90$\% level
in our work).

The steepening in $\alpha$ that we observe at faint magnitudes is limited
in significance by small number statistics, and almost certainly $\alpha$ is steeper than our fits would indicate, because of magnitude dependent 
incompleteness. Clearly much further observational work is needed  
to improve the completeness of the census of Local Group members at faint absolute magnitudes.

\acknowledgments

C.J.P. acknowledges the hospitality of the South African Astronomical
Observatory, where part of this work was completed, and the financial
support of the Natural Sciences and Engineering Research Council of
Canada. The authors are grateful to George Jacoby and Taft Armandroff
for pointing out an error in an earlier version of Fig. 2.

\clearpage

\begin{deluxetable}{llllrrrrr}
\scriptsize
\tablenum{1}
\tablewidth{0pt}
\tablecaption{Derived Properties of Probable Local Group Galaxies}
\tablehead{\colhead{Name} & \colhead{Alias} & \colhead{DDO Type} &
  \colhead{(m-M)$_0$} & \colhead{M$_V$} & \colhead{$\ell$} & 
  \colhead{$b$} & \colhead{D[kpc]} & \colhead{cos$\theta$}}
\startdata
M 31      &N 224   & Sb I-II     &24.4  &-21.2  &121.17 &-21.57 & 760  & 0.88 \\
Milky Way &Galaxy  &S(B)bc I-II: &14.5  &-20.9:  &000.00 & 00.00 &   8  &-0.15 \\
M 33      &N 598   & Sc II-III   &24.5  &-18.9  &133.61 &-31.33 & 795  & 0.73 \\
LMC       &...     & Ir III-IV   &18.5  &-18.5  &280.19 &-33.29 &  50  &-0.80 \\
SMC       &...     &Ir IV/IV-V   &18.85 &-17.1  &302.81 &-44.33 &  59  &-0.61 \\
M 32      &NGC 221 & E2          &24.4  &-16.5  &121.15 &-21.98 & 760  & 0.88 \\
NGC 205   &...     & Sph         &24.4  &-16.4  &120.72 &-21.14 & 760  & 0.88 \\
IC 10     &...     & Ir IV:      &24.1  &-16.3  &118.97 &-03.34 & 660  & 0.94 \\
NGC 6822  &...     & Ir IV-V     &23.5  &-16.0  &025.34 &-18.39 & 500  & 0.29 \\
NGC 185   &...     & Sph         &24.1  &-15.6  &120.79 &-14.48 & 660  & 0.91 \\
IC 1613   &...     & Ir V        &23.3  &-15.3  &129.73 &-60.56 & 725  & 0.47 \\
NGC 147   &...     & Sph         &24.1  &-15.1  &119.82 &-14.25 & 660  & 0.92 \\
WLM       &DDO 221 & Ir IV-V     &24.85 &-14.4  &075.85 &-73.63 & 925  & 0.32 \\
Sagittarius& ...    & dSph(t)    &17.0  &-13.8::&005.61 &-14.09 &  24  &-0.04 \\
Fornax    &...     & dSph        &20.7  &-13.1  &237.24 &-65.66 & 138  &-0.25 \\
Pegasus   &DDO 216 & Ir V        &24.4  &-12.3  &094.77 &-43.55 & 760  & 0.76 \\
Leo I     &Regulus  & dSph       &22.0  &-11.9  &225.98 &+49.11 & 250  &-0.44 \\
And I     &...     & dSph        &24.55 &-11.8  &121.69 &-24.85 & 810  & 0.86 \\
And II    &...     & dSph        &24.2  &-11.8  &128.91 &-29.15 & 700  & 0.78 \\
Leo A     &DDO 69  & Ir V        &24.2  &-11.5  &196.90 &+52.41 & 690  &-0.14 \\
Aquarius* &DDO 210 & Ir V        &25.05 &-11.3  &034.04 &-31.35 &1025  & 0.40 \\
SagDIG*   &...     & Ir V        &25.7: &-10.7: &021.13 &-16.23 &1300: & 0.22 \\
Pegasus II&And VI  & dSph        &24.45 &-10.6  &106.01 &-36.30 & 830  & 0.83 \\ 
Pisces    &LGS 3   & dIr/dSph    &24.55 &-10.4  &126.77 &-40.88 & 810  & 0.71 \\
And III   &...     & dSph        &24.4  &-10.2  &119.31 &-26.25 & 760  & 0.86 \\
And V     &...     & dSph        &24.55 &-10.2  &126.22 &-15.12 & 810  & 0.87 \\
Leo II    &...     & dSph        &21.6  &-10.1  &220.14 &+67.23 & 210  &-0.26 \\
Phoenix   &...     & dIr/dSph    &23.0  & -9.8  &272.19 &-68.95 & 395  &-0.30 \\
Sculptor  &...     & dSph        &19.7  & -9.8  &287.69 &-83.16 &  87  &-0.06 \\
Tucana    &...     & dSph        &24.7  & -9.6  &322.91 &-47.37 & 870  &-0.44 \\
Cassiopeia&And VII & dSph        &24.2  & -9.5  &109.46 &-09.95 & 690  & 0.98 \\ 
Sextans   &...     & dSph        &19.7  & -9.5  &243.50 &+42.27 &  86  &-0.65 \\
Carina    &...     & dSph        &20.0  & -9.4  &260.11 &-22.22 & 100  &-0.85 \\
Draco     &...     & dSph        &19.5  & -8.6  &086.37 &+34.71 &  79  & 0.77 \\
Ursa Minor&...     & dSph        &19.0  & -8.5  &104.88 &+44.90 &  63  & 0.66 \\
\enddata
\tablecomments{Colons denote uncertain values.}
\tablenotetext{*} {Membership in Local Group not yet firmly established.}
\end{deluxetable}

\begin{deluxetable}{ccrrrr}
\scriptsize
\tablenum{2}
\tablewidth{0pt}
\tablecaption{Maximum Likelihood Fits to Local Group Data}
\tablehead{\colhead{M$_V$(br)} & \colhead{M$_V$(ft)} & \colhead{$\alpha$} &
  \colhead{$\alpha$(95\%)} & \colhead{$\alpha$(99\%)} & 
  \colhead{\hfil P($\alpha<-1.3$)} }
\startdata
\multicolumn{6}{l}{(a) All morphological types} \\
 -22 &  -8 &  -1.07 $\pm$ 0.05 &  -0.98 &  -0.94 &  0.00\% \\
 -22 & -10 &  -1.09 $\pm$ 0.07 &  -0.96 &  -0.91 &  0.50\% \\
 -22 & -11 &  -1.02 $\pm$ 0.08 &  -0.88 &  -0.83 &  0.18\% \\
 -22 & -12 &  -0.91 $\pm$ 0.10 &  -0.74 &  -0.66 &  0.02\% \\
 -18 &  -8 &  -1.10 $\pm$ 0.07 &  -0.99 &  -0.93 &  0.40\% \\
 -18 & -10 &  -1.17 $\pm$ 0.10 &  -0.99 &  -0.94 & 11.31\% \\
 -18 & -11 &  -1.09 $\pm$ 0.13 &  -0.87 &  -0.77 &  6.52\% \\
 -18 & -12 &  -0.92 $\pm$ 0.18 &  -0.61 &  -0.45 &  2.13\% \\
 -16 &  -8 &  -1.09 $\pm$ 0.09 &  -0.94 &  -0.88 &  1.82\% \\
 -16 & -10 &  -1.21 $\pm$ 0.14 &  -0.95 &  -0.87 & 30.11\% \\
 -16 & -11 &  -1.06 $\pm$ 0.20 &  -0.72 &  -0.57 & 15.15\% \\
 -16 & -12 &  -0.62 $\pm$ 0.35 &   0.05 &   0.32 &  2.38\% \\
 -15 &  -8 &  -1.18 $\pm$ 0.11 &  -0.98 &  -0.92 & 17.93\% \\
 -15 & -10 &  -1.50 $\pm$ 0.21 &  -1.14 &  -1.01 & 85.87\% \\
 -15 & -11 &  -1.48 $\pm$ 0.33 &  -0.94 &  -0.71 & 73.27\% \\
\multicolumn{6}{l}{(b) Sph/dSph data only} \\
 -17 &  -9 &  -1.33 $\pm$ 0.12 &  -1.13 &  -1.05 & 62.53\% \\
 -17 & -10 &  -1.31 $\pm$ 0.16 &  -1.04 &  -0.94 & 58.55\% \\
 -17 & -11 &  -1.11 $\pm$ 0.22 &  -0.73 &  -0.57 & 23.66\% \\
 -15 &  -9 &  -1.52 $\pm$ 0.18 &  -1.24 &  -1.12 & 90.72\% \\
 -15 & -10 &  -1.66 $\pm$ 0.29 &  -1.23 &  -1.05 & 92.01\% \\
 -15 & -11 &  -1.44 $\pm$ 0.46 &  -0.72 &  -0.42 & 66.92\% \\
\multicolumn{6}{l}{(c) Ir/dIr data only} \\
 -16 &  -9 &  -1.02 $\pm$ 0.17 &  -0.70 &  -0.58 &  7.90\% \\
 -16 & -10 &  -1.09 $\pm$ 0.22 &  -0.72 &  -0.56 & 20.69\% \\
 -16 & -11 &  -1.01 $\pm$ 0.31 &  -0.48 &  -0.26 & 20.49\% \\
 -15 & -10 &  -1.40 $\pm$ 0.33 &  -0.90 &  -0.66 & 66.20\% \\
\enddata
\tablecomments{M$_V$(br) and M$_V$(ft) give the range of absolute magnitude
over which the fit was done. The error in $\alpha$ corresponds to a 1$\sigma$
error. $\alpha$(95\%) and $\alpha$(99\%) are the values for which
the probability is 95\% (99\%) that the true value of $\alpha$ lies below
the tabulated values. P($\alpha<-1.3$) is the probability that that 
$\alpha < -1.3$.}
\end{deluxetable}

\clearpage
\begin{center}
FIGURE CAPTIONS
\end{center}

\figcaption[Fig1.ps]{Histogram of the luminosity distribution of Local
Group members, with absolute magnitude data taken from Table 1. The upper
panel gives the luminosity distribution for all Local Group members,
and the lower panels show the luminosity distributions for Ir/dIr and Sph/dSph
morphological types. Four galaxies appear in the top panel but not in the lower
panels: the spirals M31, the Milky Way, and M33; and the elliptical M32. The
galaxies Pisces and Phoenix (dIr/dSph) are counted with weight 0.5 in each
of the lower two panels.}

\figcaption[Fig2.ps]{Cumulative distribution of absolute magnitudes of
Local Group members ({\it solid line}), compared with cumulative Schechter
functions ({\it dotted lines}). The Schechter functions are computed
with M$_V^*$=$-$20 and five different $\alpha$ values (--0.9 to --1.3 in
steps of 0.1). The cumulative functions are all normalized at M$_V=-10$
because it seems likely that the data are incomplete below this level.
The {\it dashed line} shows the empirical luminosity function (in
cumulative form) of Trentham (1998$c$).}

\clearpage

\begin{figure}
\figurenum{1}
\plotone{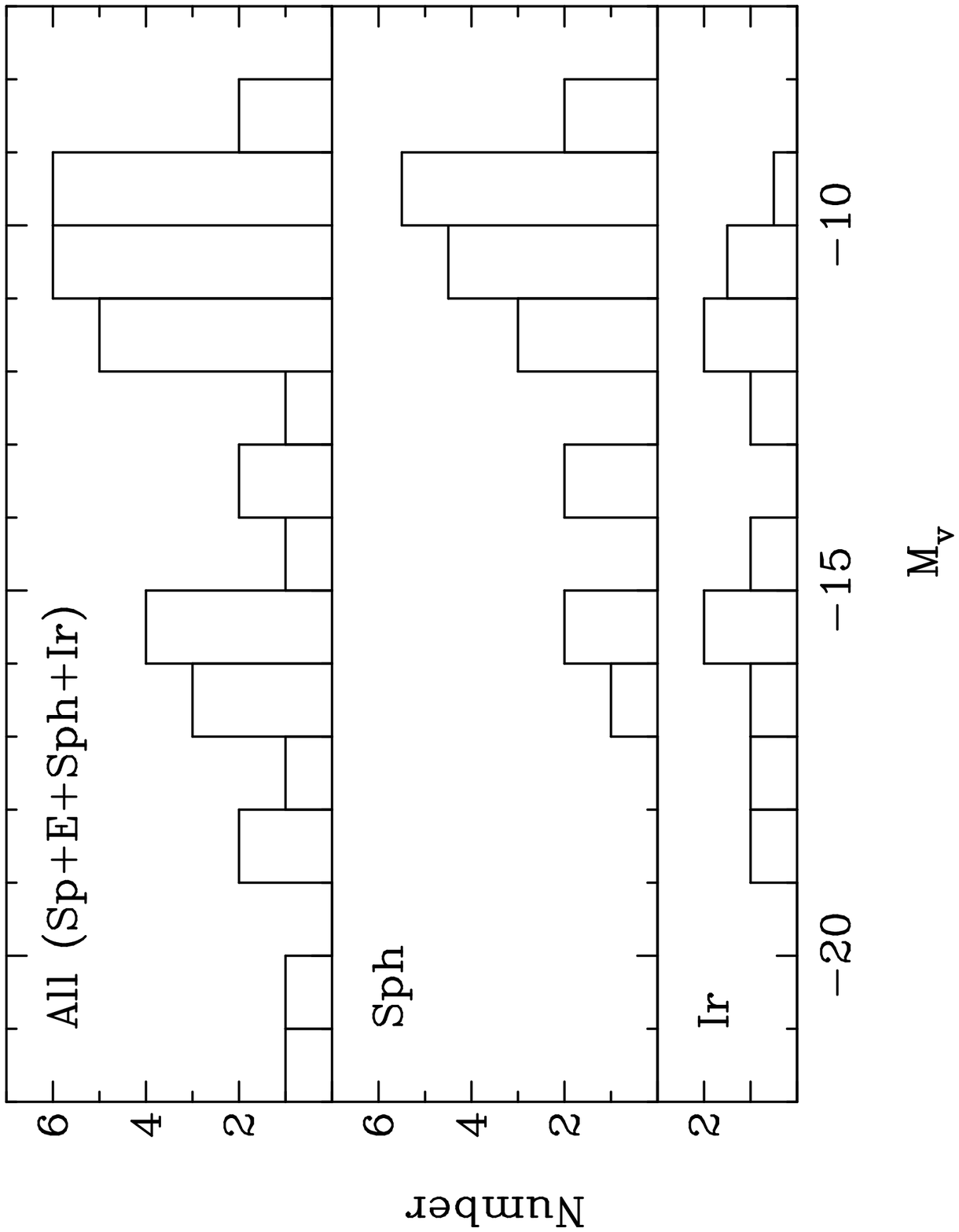}
\end{figure}

\begin{figure}
\figurenum{2}
\plotone{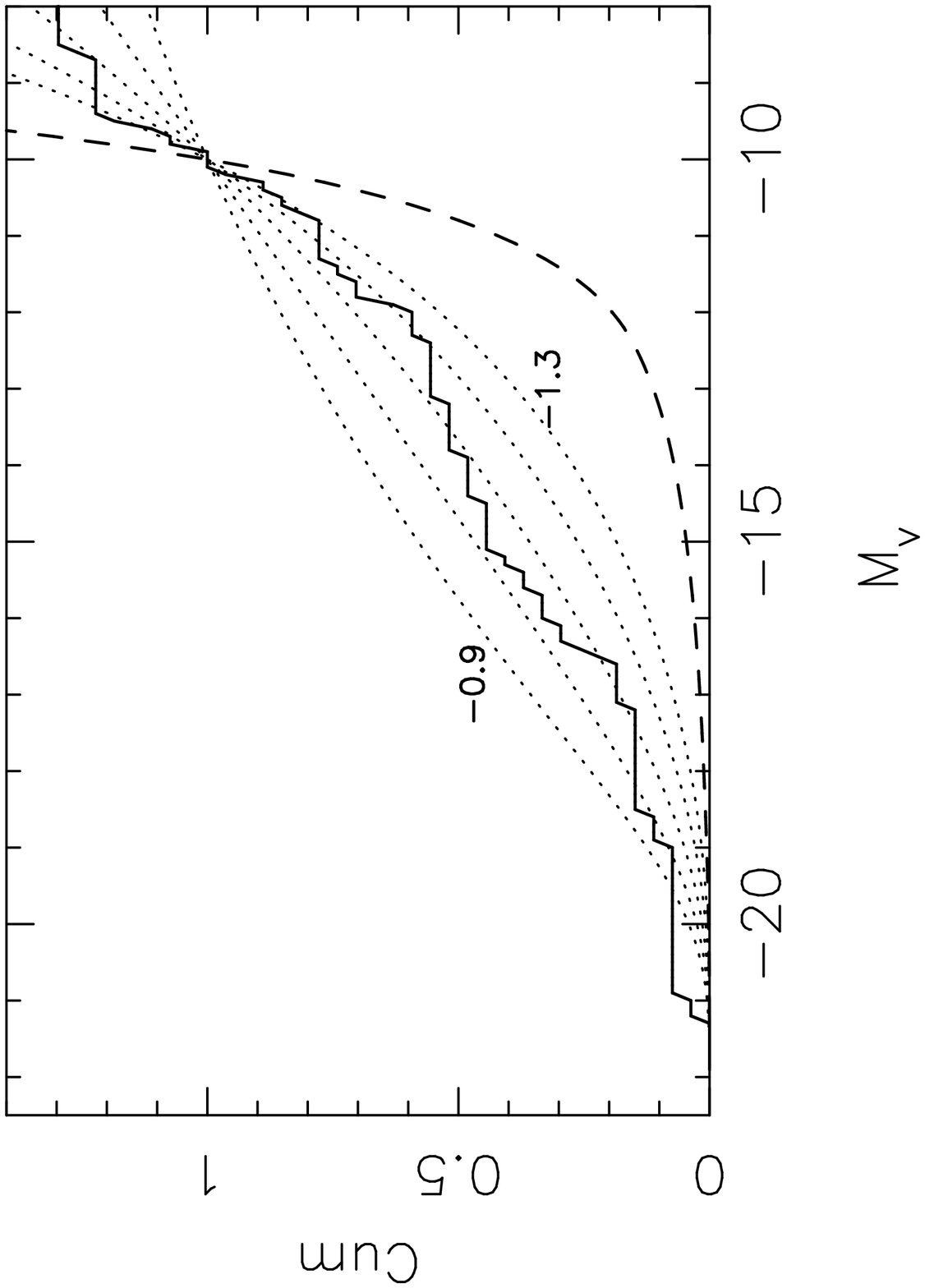}
\end{figure}


\begin{references}
\reference{} Babul, A., \& Ferguson, H. 1996, \apj, 458, 100
\reference{} Binggeli, B. Sandage, A., \& Tammann, G. A. 1988, \araa, 26, 509
\reference{} Bothun, G. D., Impey, C. D., \& Malin, D. F. 1991, \apj, 376, 404
\reference{} Bromley, B.C., Press, W.H., Lin, H., \& Kirshner, R. P. 1998, 
  \apj, in press (astro-ph 9711227)
\reference{} Braun, R., \& Burton, W. B. 1999, \aap, in press (astro-ph 9810433)
\reference{} Caldwell, N., Armandroff, T. E., Da Costa, G. S., \& Seitzer, P. 1998, AJ, 115, 535
\reference{} Caldwell, N., \& Schommer, R. 1988, in ASP Conf. Ser. 4, The Extragalactic Distance Scale, ed. S. van den Bergh \& C. J. Pritchet (San Francisco: ASP), 77
\reference{} C\^ot\'e, S., Broadhurst, T., Loveday, J., \& Kolind, S. 1998,
  in IAU Colloquium  171, The Low Surface Brightness Universe, ed.
  J.I. Davies \etal (San Francisco: ASP), in press (astro-ph 9810470)
\reference{} Courteau, S., \& van den Bergh, S. 1999, \aj, in press
\reference{} Dohm-Palmer, R. C., Skillman, E. D., Gallagher, J., Tolstoy, E., Mateo, M., Dufour, R. J., Saha, A., Hoessel, J., \& Chiosi, C. 1998, \aj, 116, 1227
\reference{} Felten, J. E. 1985, Comm. Astrophys., 11, 53
\reference{} Ferguson, H., \& Sandage, A. R. 1991, \aj, 101, 765
\reference{} Frenk, C., Evrard, A. E., White, S. D. M., \& Summers, F. J. 1996, \apj, 472, 460
\reference{} Hubble, E. 1936, The Realm of the Nebulae (New Haven:Yale University Press)
\reference{} Hunsberger, S. D., Charlton, J. C., \& Zaritsky, D. 1998, \apj, 
  505, 536
\reference{} Impey, C. D., Bothun, G. D., \& Malin, D. F. 1988, \apj, 330, 634
\reference{} Irwin, M. J., in Dwarf Galaxies, ed. G. Meylan \& P. Prugniel (Garching:ESO), 27
\reference{} Jerjen, H., Freeman, K. C., \& Binggeli, B. 1998, \aj, 116, 2873
\reference{} Karachentsev, I. D., Tikhonov, N. A., \& Sazonova, L. N. 1994, \aplett, 20, 84
\reference{} Kauffmann, G., Colberg, J. M., Diaferio, A., \& White, S. D. M. 1998, \mnras, submitted (astro-ph 9805283)
\reference{} Krismer, M., Tully, R. B., \& Gioia, I. M. 1995, \aj, 110, 1584
\reference{} Lin, H., Kirshner, R. P., Shectman, S. A., Landy, S. D., 
  Oemler, A., Tucker, D. L., \& Schechter, P. L. 1996, \apj, 464, 60
\reference{} L\'opez-Cruz, O., Yee, H. K.-C., Brown, J. P., Jones, C.,
  \& Forman, W. 1997, \apj, 475, L97
\reference{} Loveday, J. 1997, \apj, 489, 29
\reference{} Loveday, J., Peterson, B. A., Efstathiou, G., \& Maddox, S. J. 
  1992, \apj, 390, 338
\reference{} Marzke, R. O., da Costa, L. N., Pellegrini, P. S., Willmer,
  C. N. A., \& Geller, M. J. 1998, \apj,
  in press (astro-ph 9805218)
\reference{} Marzke, R. O., Geller, M. J., Huchra, J. P., \& Corwin, H. G. 1994$a$, \aj, 108, 2
\reference{} Marzke, R. O., Huchra, J. P., \& Geller, M. J. 1994$b$, \apj, 428, 43
\reference{} Mateo, M. 1998, \araa,  36, 435
\reference{} Muriel, H., Valotto, C. A., \& Lambas, D. G. 1998, \apj, in press
  (astro-ph 9804300)
\reference{} Phillipps, S., Parker, Q. A., Schwartzenberg, \& Jones, J. B. 
  1998$a$, \apjl, in press (astro-ph 9712027)
\reference{} Phillipps, S., Jones, J. B., Smith, R. B., Couch, W. J.,
  \& Driver, S. P.  1998$b$, in IAU Colloquium  171, The Low Surface 
  Brightness Universe, ed. J.I. Davies \etal (San Francisco: ASP), 
  in press (astro-ph 9812229)
\reference{} Press, W. H., \& Schechter, P. L. 1974, \apj, 187, 425
\reference{} Schechter, P. 1976, \apj, 203, 297
\reference{} Schneider, S. E., Spitzak, J. G., \& Rosenberg, J. L. 1998, \apjl, 507, L9
\reference{} Tolstoy, E., Saha, A., Hoessel, J. G., \& Danielson, G. E. 1995, AJ, 109, 579
\reference{} Trentham, N. 1998$a$, in Dwarf Galaxies and Cosmology, ed. T. X. Thuan \etal (Gif-sur-Yvette: Editions Fronti\'eres), in press (astro-ph 9804013)
\reference{} Trentham, N. 1998$b$, \mnras, in press (astro-ph 9708189)
\reference{} Trentham, N. 1998$c$, \mnras, 294, 193
\reference{} Tully, R. B. 1988, \aj, 96, 73
\reference{} van den Bergh, S. 1962, Zs. f. Astrophysik, 55, 21
\reference{} van den Bergh, S. 1992, \aap, 264, 75
\reference{} van den Bergh, S. 1994, \aj, 107, 1328
\reference{} van den Bergh, S. 1999, \apjl, in press
\reference{} van den Bergh, S. 2000, The Galaxies of the Local Group (Cambridge: Cambridge University Press)
\reference{} van Driel, W., Kraan-Korteweg, R. C., Binggeli, B., \&
Huchtmeirer, W. K. 1998, \aaps, 127, 397
\reference{} Zijlstra, A. A., \& Minniti, D. 1999, \aj, in press (astro-ph 9812330)

\end{references}
\end{document}